\documentclass[twocolumn,amsmath,amssymb]{revtex4}

\usepackage{graphicx}
\usepackage{dcolumn}
\usepackage{bm}

\let\citep=\cite

\begin{document}

\title{Effect of pooling samples on the efficiency of comparative
studies using microarrays}
\author{\small Shu-Dong Zhang and Timothy W. Gant\\
\small MRC Toxicology Unit, Hodgkin Building, Lancaster Road,
University of Leicester, Leicester, UK }

\begin{abstract}
Many biomedical experiments are carried out by pooling individual
biological samples. However, pooling samples can potentially hide
biological variance and give false confidence concerning the data
significance. In the context of microarray experiments for
detecting differentially expressed genes, recent publications have
addressed the problem of the efficiency of sample-pooling, and
some approximate formulas were provided for the power and sample
size calculations. It is desirable to have exact formulas for
these calculations and have the approximate results checked
against the exact ones. We show that the difference between the
approximate and exact results can be large.

In this study, we have characterized quantitatively the effect of
pooling samples on the efficiency of microarray experiments for
the detection of differential gene expression between two classes.
We present exact formulas for calculating the power of microarray
experimental designs involving sample pooling and technical
replications. The formulas can be used to determine the total
numbers of arrays and biological subjects required in an
experiment to achieve the desired power at a given significance
level. The conditions under which pooled design becomes preferable
to non-pooled design can then be derived given the unit cost
associated with a microarray and that with a biological subject.
This paper thus serves to provide guidance on sample pooling and
cost effectiveness. The formulation in this paper is outlined in
the context of performing microarray comparative studies, but its
applicability is not limited to microarray experiments. It is also
applicable to a wide range of biomedical comparative studies where
sample pooling may be involved.

A Java Webstart application can be accessed at
http://wads.le.ac.uk/htox/WadsSoftware

\noindent/MrcStats/SCal4Poolings.jnlp

\end{abstract}

\maketitle

\section{Introduction}
\label{sec-introduction}

Pooling samples in biomedical studies has now become a frequent
practice among many researchers. For example, more than $15\%$ of
the data sets deposited in the Gene Expression Omnibus Database
involve pooled RNA samples\citep{Kendziorski-etal-PNAS2005}. The
practice of pooling biological samples though is not a new
phenomenon, as it can be traced back at least to 1940s
\citep{Dorfman1943} and has been used in different application
areas \citep{Gastwirth2000}, e.g., for the detection of certain
medical conditions and estimation of prevalence in a population.
In the context of detecting differential gene expressions using
microarrays, divergent views on the wisdom of pooling samples can
be found in the literature
\citep{Agrawal-etal-JNatCancerInst2002,Affymetrix2004,
Shih-etal-Bioinformatics2004, Churchill&Oliver-NatureGenetics2001,
Peng-etal-BMCBioinformatics2003,Jolly-etal-PhysiolGenomics2005}.
One of the arguments supporting the practice of pooling biological
samples is that biological variation can be reduced by pooling RNA
samples in microarray
experiments\citep{Churchill&Oliver-NatureGenetics2001}. As more
carefully described by Kendziorski {\em et al}
\citep{Kendziorski-etal-PNAS2005}, pooling can reduce the effects
of biological variation, but not the biological variation itself.
Another argument in support of pooling samples in microarray
experiments is that it reduces financial cost. However, cost
reduction is meaningful only if statistical equivalence between
the pooled and the non-pooled experimental setups is maintained.
Here we address this issue and present formulas to determine the
conditions under which pooled and non-pooled designs are
statistically equivalent.

To compare experimental designs with and without sample pooling
the two designs must have something in common that can be
measured, e.g., using the same or equivalent amount of resources,
or, yielding the same level of detection power. Kendziorski {\em
et al}\citep{Kendziorski-etal-Biostatistics2003} used the width of
the 95\% confidence interval for gene expression to compare
different experimental designs with and without sample pooling.
The criterion was that the narrower the confidence interval, the
more accurate the results from the experimental design. In a
comparative study where two groups of biological subjects are
compared the common goal of the different experimental designs is
to detect a change between the two groups with a given power at a
given false positive rate, as adopted in
\citep{Shih-etal-Bioinformatics2004}. We shall use the latter
method to compare different designs. So in this work statistical
equivalence means that the designs have the same statistical power
at the same level of significance. Therefore the more appropriate
experimental design will be the one which uses less resources to
achieve this statistical equivalence.

The basic assumption underlying sample pooling is biological
averaging; that the measure of interest taken on the pool of
samples is equal to the average of the same measure taken on each
of the individual samples which contributed to the pool. For
example in the situation of a microarray experiment, if $r$
individual samples contribute equally to a pool, and the
concentrations of a gene's mRNA transcripts for the $r$ samples
are denoted by $T_i$ with $i=1, 2, \cdots, r$ indexing the
individual samples, the assumption of biological averaging says
that the concentration of this gene's mRNA transcripts in the pool
is $T=1/r\sum_{i=1}^rT_i$. However, for microarray experiment
there is some debate on whether the basic assumption of pooling
holds. Kendziorski {\em et al}
\citep{Kendziorski-etal-Biostatistics2003,Kendziorski-etal-PNAS2005}
argue that there is limited support for this assumption. Here we
do not seek to enter into this debate but rather take the
assumption of biological averaging as valid, or at least
approximately so, so that we are in a position to determine
whether pooling samples is financially beneficial or not. The
validity of biological averaging makes it possible (or easier) to
derive a neat theoretical formulation. On a practical level
though, the requirement for the validity of this assumption may
not be as stringent as a theoretical formulation does. For
instance, in \citep{Kendziorski-etal-PNAS2005} it was shown that
even when biological averaging does not hold, pooling can be
useful and inferences regarding differential gene expression are
not adversely affected by pooling.

One situation where there is little alternative but to pool
biological samples is where there is insufficient amount of RNA
from each individual biological subject to perform single
microarray hybridization. RNA amplification may be a possible way
of obtaining more RNA, but may not be practically feasible when
many individual biological subjects are involved as in the case of
\citep{Jin-etal2001}. In such a circumstance, pooling samples is
justified by the lack of alternative and will not be considered
further here. Similarly we will not consider here the case where
all the biological samples of the same group were pooled together,
and multiple technical replicate measurements were carried out on
the sample pool. This is sometimes seen in the literature
\citep{Muckenthaler-etal2003}, but such an experimental design
leaves no degree of freedom to estimate the biological variance.
Thus valid inferences about the differences between the two
populations of biological subjects under study cannot be made.
Here we only consider situations other than the above two and
where pooling may reduce the overall costs of the experiments.

\section{A general formalism}
\label{sec-formalism}

For every comparative study, there is at least one measurable
quantity which is the quantity of interest. The goal of the study
is to deduce from the data collected if there is any difference
between the means of the two populations. As measuring all the
biological subjects in two populations is rarely possible in most
situations representatives from a population are randomly selected
and measurements made on these. These are then taken to infer the
properties of the population.

Let $X$ be the measurable quantity that is being determined in the
the experiment, e.g., the expression level of a gene. In the case
of one-channel microarray, $X$ could denote the logarithm (most
commonly base 2 is used) of fluorescence intensity; or the
logarithm of the fluorescence ratio in the case of two-channel
microarray. Let $x^c_i$ denote the value of $X$ for an individual
subject $i$ in the control population (c), and $x^t_j$ that of the
individual subject $j$ in the treatment population (t). We assume
that $x^c_i$s for all individuals in the control population are
independent normally distributed with a mean $\mu_c$ and a
variance $\sigma _c^2$, denoted by $x^c_i \sim N(\mu_c,\sigma
_c^2)$ for all $i$. Similarly, $x^t_j \sim N(\mu_t,\sigma _t^2)$
for all $j$.

\subsection{A general experimental setup}
For a general experimental setup individual subjects from both
populations are randomly selected and tissue samples collected
from each. Tissue sample pools are made by pooling a given number
$r$ of randomly selected tissue samples (of the same population)
together. Note that to make $n$ pools we need to have selected
$nr$ individual subjects from the population. $m$ measurements are
then made on each pool of tissue samples. So $m$ is the number of
technical replications of measurement on each pool. Notice that by
introducing two parameters $r$ and $m$ a general and flexible
experimental setup has been created. For instance, if we set
$r=1$, the experiment would be equivalent to no pooling of tissue
samples. And if we set $m=1$ there is no technical replication.
Under the basic assumption of biological averaging, the result of
pooling $r$ tissue samples in equal proportions together is that
the value of $X$ for the pool is the average of those subjects
which formed this pool,
\begin{equation}
\tilde{x}=\frac{1}{r}\sum _{i=1}^{r}x_i.
\end{equation}
It follows that $\tilde{x} \sim N(\mu_c, \sigma_c ^2/r)$ for a
pool from the control population, or $\tilde{x} \sim N(\mu_t,
\sigma_t ^2/r)$ for a pool from the treated population. Note that
in this paper we shall only discuss pooling samples with equal
individual contributions. While pools formed by un-equal
contributions from individual samples are possible, such pooled
experimental design is generally less effective than the equal
pooling, as already shown by Peng {\em et al}
\citep{Peng-etal-BMCBioinformatics2003} with their simulated
results.

When we take a measurement on a pool $p$, the measured value is
\begin{equation}
y_{p,k}=\tilde{x}_p+\epsilon _k,
\end{equation}
where $p$ indexes pools, $k$ indexes measurements, and $\epsilon
_k$ is a random error term assumed to be independently and
normally distributed as $\epsilon _k \sim
N(0,\sigma^2_{\epsilon})$. Hereafter $\sigma^2_{\epsilon}$ will be
referred to as the technical variance, $\sigma^2_c$ the biological
variance for the control population, and $\sigma^2_t$ the
biological variance for the treatment population.

The output of the experiment are the measurements on the two sets
of pools. For the control group, we have $y^c_{p,k}$ for
$p=1,\dots, n_{c}$ and $k=1,\dots, m$. And for the treatment
group, we have $y^t_{p,k}$ for $p=1,\dots, n_{t}$ and $k=1,\dots,
m$. Here $n_{c}$ and $n_{t}$ are the numbers of pools prepared for
the control and treatment population respectively. Our task is to
infer population properties from these measured data. In
particular, we want to know whether there is any difference
between the two populations means $\mu_c$ and $\mu_t$. It can be
shown that
\begin{equation}
\overline{Y}^c=\frac{1}{mn_{c}}\sum_{p=1}^{n_c}\sum_{k=1}^my^c_{p,k}
\label{eq-Y^c}
\end{equation}
is an unbiased estimator of $\mu_c$, with a variance
\begin{equation}
\frac{1}{n_c}\left(\frac{\sigma^2_c}{r}+\frac{\sigma^2_{\epsilon}}{m}\right),
\end{equation}
and similarly,
\begin{equation}
\overline{Y}^t=\frac{1}{mn_{t}}\sum_{p=1}^{n_t}\sum_{k=1}^my^t_{p,k}
\label{eq-Y^t}
\end{equation}
is an unbiased estimator of $\mu_t$, with a variance
\begin{equation}
\frac{1}{n_t}\left(\frac{\sigma^2_t}{r}+\frac{\sigma^2_{\epsilon}}{m}\right).
\end{equation}
If we make an additional assumption that the variances for the two
populations of biological subjects are the same, i.e.,
$\sigma^2_c=\sigma^2_t=\sigma^2$, then the difference between
Eqs.(\ref{eq-Y^t}) and  (\ref{eq-Y^c}),
$D=\overline{Y}^t-\overline{Y}^c$, is an unbiased estimator of
$\mu=\mu_t-\mu_c$ with a variance
\begin{equation}
\sigma^2_D=\left(\frac{1}{n_c}+\frac{1}{n_t}\right)
\left(\frac{\sigma^2}{r}+\frac{\sigma^2_{\epsilon}}{m}\right).
\label{eq-var(D)}
\end{equation}
The factor $(\sigma^2/r+\sigma^2_{\epsilon}/m)$ in
Eq.(\ref{eq-var(D)}) can be estimated without bias by

\begin{eqnarray}
s^2_p=\frac{1}{n_c+n_t-2}
\sum_{p=1}^{n_c}\left(\frac{1}{m}\sum_{k=1}^my^c_{p,k}-\overline{Y}^c\right)^2 \nonumber \\
+\frac{1}{n_c+n_t-2}
\sum_{p=1}^{n_t}\left(\frac{1}{m}\sum_{k=1}^my^t_{p,k}-\overline{Y}^t\right)^2.
\end{eqnarray}
It is then clear that
\begin{equation}
t=\frac{(\overline{Y}^t-\overline{Y}^c)-(\mu_t-\mu_c)}{s_p\sqrt{1/n_c+1/n_t}}
\end{equation}
follows the Student's t distribution with $n_c+n_t-2$ degrees of
freedom. In detecting a differential gene expression, we want to
test the null hypothesis $\mu_c=\mu_t$ against an alternative
hypothesis $\mu_c\neq \mu_t$. So our test statistic is
\begin{equation}
t_0=\frac{(\overline{Y}^t-\overline{Y}^c)}{s_p\sqrt{1/n_c+1/n_t}},
\label{eq-t_0}
\end{equation}
and there are no unknowns in Eq.(\ref{eq-t_0}). Note that $t_0$
can be seen as a generalized two-sample-t-test statistic, which
reduces to the statistic of the traditional two-sample t test with
equal variance when we set the parameters $r=1$ (no pooling of
tissue samples) and $m=1$ (no technical replication of
measurements). In Ref. \citep{Shih-etal-Bioinformatics2004},
\citeauthor{Shih-etal-Bioinformatics2004} arrived at two separate
statistics, one for non-pooled design, the other for pooled
design. The $t_0$ defined by Eq.(\ref{eq-t_0}) is in more general
form, setting $r=1$ and $m=1$ in Eq.(\ref{eq-t_0}) recovers Shih
{\em et al}'s statistic for non-pooled design; while setting $r>1$
and $m=1$ recovers Shih {\em et al}'s statistic for pooled design.
Note that $m$ does not need to equal 1. Here by incorporating two
additional parameters $r$ and $m$, the statistic $t_0$ can deal
with situations where there are pooled tissue samples and multiple
technical replications.

\subsection{Criteria of significance}
As with any statistical test we need to specify a threshold
p-value $P_{th}$ to claim significant results in the test. When
all the other parameters are given, setting $P_{th}$ is equivalent
to setting a threshold, say $|\xi|$, for the statistics $t_0$
defined in Eq.(\ref{eq-t_0}). With this threshold t-value, our
criteria for claiming a significant test is as follows: If
$t_0>|\xi|$, we declare that $\mu_t-\mu_c>0$; if $t_0<-|\xi|$, it
is claimed as $\mu_t-\mu_c<0$. So the rate at which false positive
claims are made is
\begin{eqnarray}
P_{th}=\int _{-\infty}^{-|\xi|}\rho _{n_c+n_t-2}(t_0)dt_0
+\int_{|\xi|}^{\infty}\rho _{n_c+n_t-2}(t_0)dt_0 \nonumber \\
=2\int _{-\infty}^{-|\xi|}\rho _{n_c+n_t-2}(t_0)dt_0
=2T_{n_c+n_t-2}(-|\xi|), \label{eq-P_th}
\end{eqnarray}
where $\rho_{n_c+n_t-2}(.)$ is the probability density function
(PDF) of the Student's t distribution with $n_c+n_t-2$ degrees of
freedom, and $T_{n_c+n_t-2}(.)$ is the corresponding cumulative
probability distribution function (CDF). It is therefore apparent
that the threshold t-value $|\xi|$ can be obtained by solving the
equation $2T_{n_c+n_t-2}(-|\xi|)=P_{th}$ with a given false
positive rate $P_{th}$.

\section{Power function}
\label{sec-power-function} In
\citep{Zhang&Gant-Bioinformatics2004} we presented a power
function for a new statistical t test (hereafter referred to as
"two-labelling t test") in the context of using two-color
microarrays to detect differential gene expression. Following
similar steps we can derive the power function for the generalized
two-sample t test presented in this paper, which reads

\begin{eqnarray}
S=\int _0^{\infty}p_{n_c+n_t-2}(Y)\Phi\left[{-|\xi|\sqrt{Y}\over
\sqrt{n_c+n_t-2}} +\frac{|\mu|}{\sigma_D} \right]dY,
 \label{eq-S}
\end{eqnarray}
where $p_{n_c+n_t-2}(Y)$ is the PDF for the $\chi ^2$ distribution
with $n_c+n_t-2$ degrees of freedom, and $\Phi(.)$ is the CDF for
the standard normal distribution. The rate $S$ at which a true
difference between $\mu_t$ and $\mu_c$ can be successfully
detected is a function of $n_c$, $n_t$, $|\mu|/\sigma _D$, and
$|\xi|$. With $\sigma_D$ given by the square root of
Eq.(\ref{eq-var(D)}), and $|\xi|$ determined by solving
Eq.(\ref{eq-P_th}) at a given false positive rate $P_{th}$, $S$
is, eventually, a function of $P_{th}$, $n_c$, $n_t$, and
$|\mu|/\sigma _D$.

A few points are worth noting here.

1. The two-labelling t test presented in
\citep{Zhang&Gant-Bioinformatics2004} was designed to deal with
systematic labelling biases generated during microarray
experimentation. The t test presented in this paper, however,
assumes no systematic data biases. In the case of two-color
microarrays this requires a common reference design. In such an
experimental design the labelling biases cancel themselves out in
the calculation of the test statistic.

2. In \citep{Zhang&Gant-Bioinformatics2004}, the biological
variances of the two populations under comparison do not have to
be the same, that is, we did not assume $\sigma^2_c=\sigma^2_t$.
For the t test in this paper, we have made an additional
assumption that $\sigma^2_c=\sigma^2_t$. Relaxing this requirement
was possible, as in the case of the traditional two-sample t test
with unequal variance \citep{Brownlee1965}, but an exact power
function could not be readily obtained.

3. The exact power function obtained in this paper allows
evaluation of the effects of pooling biological samples and the
effects of taking multiple technical measurements, thus giving
researchers quantitative guidance on the practice of pooling
samples.

4. By setting the parameters $r=1$ and $m=1$, an exact power
function is provided for the traditional two-sample t test with
equal variance.

\section{Results}

We have implemented the computation of the power function $S$ of
Eq.(\ref{eq-S}) as a Java application, which can be accessed at
the URL given in the abstract. Here we apply this to microarray
comparative studies for finding differentially expressed genes,
and investigate the effect of pooling RNA samples in the
experiments. We also compare our exact results with some
approximate results presented by other authors
\citep{Shih-etal-Bioinformatics2004} to demonstrate why an exact
formula is desirable.

\subsection{Comparison with approximate results}
Based on their approximate formulas,
\citeauthor{Shih-etal-Bioinformatics2004} considered two scenarios
to compare the number of biological subjects and number of
microarrays in the non-pooled and pooled
designs\citep{Shih-etal-Bioinformatics2004}. Here we give exact
results for the two scenarios to show the difference to the
approximate results. In the first scenario, we consider that the
common biological variance of the two populations is $\sigma
^2=0.05$, and the technical variance $\sigma_{\epsilon}^2=0.0125$,
which gives the biological-to-technical variance ratio $\lambda
=\sigma^2/\sigma_{\epsilon}^2=4$. The preset target of the
experiment in this scenario is that the false positive rate being
controlled at $P_{th}=0.001$ and the power being no less than
$S=0.95$ to detect a two-fold differential gene expression, which
corresponds to $\mu=1$ with base 2 logarithm
\citep{Shih-etal-Bioinformatics2004}. In Table
\ref{tab-scenario1}, we present results for different pooling
parameter $r$. It can be seen from the first panel of this Table
that in order to hit the preset target, the non-pooled design
($r=1$) requires at least $12$ biological subjects divided evenly
to the two populations, i.e., $6$ from each of the two
populations. Having $7$ subjects from one population and $5$
subjects from the other is insufficient to achieve the target of
$95\%$ detection power. The effects of other levels of pooling on
the detection power are also shown in Table \ref{tab-scenario1}.
The minimum number of biological subjects ($N_s$) and microarrays
($N_m$) that meet the preset targets are highlighted with bold
fonts. It is clear that as the level of pooling is increased (with
increasing $r$), the number of microarrays $N_m$ can be reduced,
but the number of biological subjects $N_s$ has to be increased.
For example, in order to reduce the number of arrays from $12$
(Table \ref{tab-scenario1}, first panel) to $8$ (Table
\ref{tab-scenario1}, fourth panel), the number of biological
subjects to form the pools must be increased from $12$ to $40$.

\begin{figure}
\centerline{\includegraphics[width=8.5cm,height=6.0cm,angle=0]{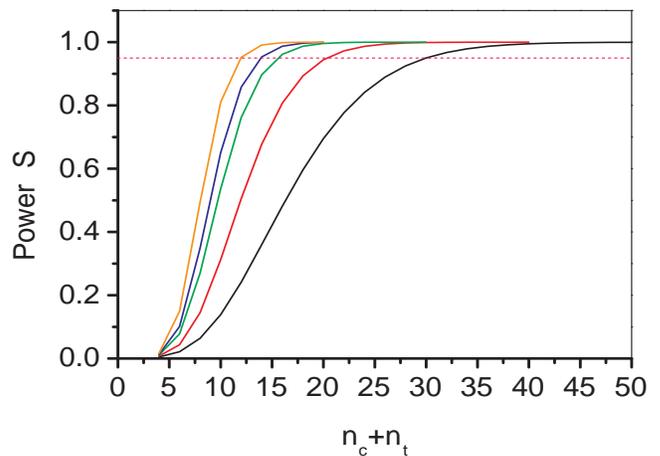}}
\caption{The power $S$ as a function of the total number of pools
$n_c+n_t$. The parameters used are for the second scenario $\sigma
^2=0.2$, $\sigma_{\epsilon}^2=0.05$, $\lambda
=\sigma^2/\sigma_{\epsilon}^2=4$, $P_{th}=0.001$, $\mu=1$, and
$m=1$. The five solid curves correspond to different levels of
pooling, from right to left, $r=1$, $r=2$, $r=4$, $r=6$, and
$r=15$ respectively. The dashed line indicates the $95\%$ power,
the intersections of which with the power curves specify the total
numbers of pools (assuming $n_c=n_t$) needed to achieve the target
power. The total number of biological subjects and the total
number of arrays can then be calculated simply by
$N_s=r(n_c+n_t)$, and $N_m=m(n_c+n_t)$ respectively.}
\label{fig-Scurve}
\end{figure}

For the second scenario we consider the case $\sigma ^2=0.2$,
$\sigma_{\epsilon}^2=0.05$, which gives $\lambda
=\sigma^2/\sigma_{\epsilon}^2=4$. Again the preset targets are to
detect a true differential expression $\mu=1$ with no less than
$95\%$ power while the false positive rate is set at
$P_{th}=0.001$. Using these parameters, the power $S$ as a
function of $n_c+n_t$ is plotted in Fig.\ref{fig-Scurve} for
different levels of sample-pooling. For the non-pooled design
($r=1$), $N_s=30$ total biological subjects and $N_m=30$ arrays
are required to hit the preset targets. Similar to the first
scenario, as the level of pooling is increased, the number of
arrays $N_m$ is reduced while the number of subjects increased to
meet the preset targets.

In Table \ref{tab-exact-vs-approx}, we summarize our exact results
and the approximate results of
\citep{Shih-etal-Bioinformatics2004}. It can be seen that the
difference between the two can be very large, indicating the need
for exact results.  For example, in the first scenario when
$N_m=8$ the approximate result of
\citep{Shih-etal-Bioinformatics2004} predicts that a minimum of
$21$ biological subjects are required. In practice $24$ subjects
are required as $24$ is the minimum number larger than $21$ and
divisible by $8$. However this experiment setup ($24$ subjects
forming $8$ pools, $8$ microarrays) will only give a detection
power of $90\%$. To meet the target power of $95\%$, $40$
biological subjects are actually required by our exact result. If
an experiment with $N_m=7$ microarrays is planned,
\citeauthor{Shih-etal-Bioinformatics2004} predicts that $37$
subjects are required\citep{Shih-etal-Bioinformatics2004}, but in
fact $126$ subjects must be used to achieve the target. Generally,
the approximate formulas of \citep{Shih-etal-Bioinformatics2004}
are too optimistic in assessing the benefits of pooling samples
and reducing the number of microarrays, because they underestimate
the number of biological subjects required.

\subsection{Cost analysis}
Depending on the material costs involved in the biological
subjects and microarrays, the conditions where pooling samples
becomes beneficial may be different from lab to lab. Here we show
examples to determine these conditions. Denoting the cost
associated with each biological subject as $C_s$ (including
materials and labor etc) and the cost associated with a microarray
as $C_m$, then the total costs for an experiment in microarray
comparative study is $C_T=N_sC_s+N_mC_m$. Taking the first
scenario as an example, the total cost of a non-pooled design to
achieve our preset targets is
$$C_T(r=1)=12C_s+12C_m,$$
and the total cost for pooled design with $r=2$ is
$$C_T(r=2)=20C_s+10C_m.$$
Therefore in order that the pooled design with $r=2$ is beneficial
we must have

\begin{equation}
C_T(r=2) \le C_T(r=1),
\end{equation}
which requires that $C_m \ge 4C_s$. Put another way, only when the
cost associated with one microarray $C_m$ is more than $4$ times
the cost of a subject $C_s$, does the pooling design with $r=2$
become preferable to the non-pooled design. Similarly a higher
level of pooling with $r=3$ becomes  preferable to $r=2$ only when
$C_m \ge 7C_s$. Furthermore the conditions for increasing the
level of pooling from $r=3$ to $r=5$ are $C_m \ge 13C_s$, and so
on. Table \ref{tab-exact-vs-approx} gives these conditions for
further levels of pooling.

For the first scenario using the actual cost figures given in
\citep{Shih-etal-Bioinformatics2004} where $C_s=\$230$ and
$C_m=\$300$, it can be seen that none of the pooling conditions is
met. Therefore for this laboratory pooling samples is not
recommended. However, if we use the cost figures of
\citep{Kendziorski-etal-Biostatistics2003} where $C_s=\$50$ and
$C_m=\$700$, an optimal design is a pooled design with $r=5$.

For the second scenario, it is a similar story. The cost figures
of Ref. \citep{Shih-etal-Bioinformatics2004} ($C_s=\$230$ and
$C_m=\$300$) gives $C_m=1.30C_s$, which does not satisfy any of
the pooling conditions. So again the non-pooled design with
$N_m=30$ and $N_s=30$ is recommended. On the other hand, the cost
figures of \citep{Kendziorski-etal-Biostatistics2003} ($C_s=\$50$
and $C_m=\$700$) give $C_m=14C_s$ which satisfies all the pooling
conditions in the lower panel of Table \ref{tab-exact-vs-approx}
except the last row. So in Kendziorski et al's lab, the pooled
design with $N_m=14$ and $N_s=84$ would be recommended.

\section{Discussion} \label{sec-discussion} We have in this paper
presented exact formulas for calculating the power of microarray
experimental design with different levels of pooling. These
formulas can be used to determine the conditions of statistical
equivalence between different pooling setups. As in
\citep{Kendziorski-etal-Biostatistics2003} and
\citep{Shih-etal-Bioinformatics2004}, the calculations presented
in this paper are for an individual gene, so the statistical
equivalence for different designs of pooling can be determined
with regard to one particular gene. However, microarray monitors
thousands of genes simultaneously, and the biological and
technical variances vary from gene to gene, therefore no single
result of statistical equivalence between pooled and non-pooled
designs applies equally to all genes on the array. So in practice
how would the formulations in this work be used? One possible way,
as suggested by Kendziorski {\em et al}.
\citep{Kendziorski-etal-Biostatistics2003}, is to specify the
distributions of $\sigma ^2$ and $\sigma_{\epsilon}$ and calculate
the total number of subjects and arrays that maximize the average
power across the the array. In theory, if the biological variances
and technical variances were known for all genes on the array, an
equivalence condition between pooled and non-pooled designs could
be determined for each gene individually. The overall (or say,
average) equivalence condition between pooled and non-pooled
designs could be obtained, for example, by some form of averaging
operation over all genes. An alternative and probably more
practical way is to use representative values of $\sigma ^2$ and
$\sigma_{\epsilon}$. We therefore propose that parameters for
"typical gene" be used as inputs for the power and sample size
calculations. A typical gene is a gene whose biological and
technical variance have the most probable values among the genes,
i.e., the mode of the distribution for biological and technical
variance of genes. Alternatively, the median or mean variances
across genes could be used as representative
values\citep{Shih-etal-Bioinformatics2004}.

An issue associated with microarray experiments is the problem of
multiple inferences, where a separate null hypothesis is being
tested for each gene. Given thousands of null hypotheses being
tested simultaneously, the customary significance level $\alpha
=0.05$ for declaring positive tests will surely give too many
false positives. For example, if among a total number $N=10000$ of
genes being tested, $N_0=4000$ are truly null genes (genes that
are non-differentially expressed between the two classes), the
expected number of false positive results would be $4000\times
0.05=200$, which may be too many to be acceptable. Thus a smaller
threshold p-value for declaring differentially expressed genes
should be used. Effectively controlling false positives in a
multiple testing situation such as microarray experiments is an
area which has drawn much attention in recent years due to the
wider application of microarray technology. As discussed in our
previous work in \citep{Zhang&Gant-Bioinformatics2004}, generally
speaking, all different multiple-testing adjustment methods
eventually amount to effectively setting a threshold p-value, and
then rejecting all the null hypotheses with p-value below this
threshold. The classical Bonferroni multiple-testing procedure,
which controls family-wise error rate at $\alpha$ by setting the
threshold $P_{th}=\alpha /N$, is generally regarded as being too
conservative in the microarray context. The FDR (False Discovery
Rate) idea, initially due to \citep{Benjamini&Hochberg1995} in
dealing with the multiple testing problem, has now been widely
accepted as appropriate to the microarray situation. Recently,
Efron \citep{Efron2004} extended the FDR idea by defining fdr, a
local version FDR (local false discovery rate). When planning
microarray experiments in terms of power and sample size
calculation, the FDR of \citep{Benjamini&Hochberg1995} is more
appropriate and convenient to use. There are now in the literature
a few slightly different variants of the definition of FDR
\citep{Benjamini&Hochberg1995,Storey&Tibshrirani-pnas2003,Grant-etal-Bioinformatics2005},
but in essence it is defined as the proportion of false positives
among all positive tests declared. To provide an interface between
FDR and the formulation in the previous sections, here we show
that there is a simple correspondence between controlling FDR and
specifying the traditional type I error rate and power. Suppose
that there are a total number $N$ of genes being monitored by
microarray, so there will be $N$ hypotheses being tested, one for
each gene. Suppose that a fraction $\pi_0$ of the $N$ genes are
true null genes, i.e., genes that are non-differentially expressed
between the two classes. Given the type I error rate $P_{th}$, the
expected number of false positive tests is $P_{th}N\pi_0$; Given
the power $S$, the expected number of non-null genes (truly
differentially expressed genes) that are declared positive is
$SN(1-\pi_0)$. So the FDR achieved by this setting is
\begin{equation}
\mbox{FDR}=\frac{P_{th}N\pi_0}{P_{th}N\pi_0+SN(1-\pi_0)}
=\frac{P_{th}\pi_0}{P_{th}\pi_0+S(1-\pi_0)}. \label{eq-fdr}
\end{equation}
Here $\pi_0$ is an important parameter in controlling FDR, for
which  several different methods of estimating this parameter have
been proposed
\citep{Pounds&Morris2003,Storey&Tibshrirani-pnas2003,Zhang&Gant-Bioinformatics2004}.
Especially the method we presented in
\citep{Zhang&Gant-Bioinformatics2004} is an accurate yet
computationally much simpler algorithm than the one proposed by
Storey and Tibshrirani in \citep{Storey&Tibshrirani-pnas2003}.
With the interface Eq.(\ref{eq-fdr}), FDR can be readily presented
and incorporated into the calculations.


\vspace{0.2cm}\noindent {\Large\bf Acknowledgments}

\noindent We wish to acknowledge the support of the microarray
team of the MRC Toxicology Unit particularly Reginald Davies,
JinLi Luo and Joan Riley. We also thank two anonymous reviewers
for their helpful and constructive comments.

\bibliography{poolingSamples-references}

\begin{thebibliography}{19}
\expandafter\ifx\csname natexlab\endcsname\relax\def\natexlab#1{#1}\fi
\expandafter\ifx\csname bibnamefont\endcsname\relax
  \def\bibnamefont#1{#1}\fi
\expandafter\ifx\csname bibfnamefont\endcsname\relax
  \def\bibfnamefont#1{#1}\fi
\expandafter\ifx\csname citenamefont\endcsname\relax
  \def\citenamefont#1{#1}\fi
\expandafter\ifx\csname url\endcsname\relax
  \def\url#1{\texttt{#1}}\fi
\expandafter\ifx\csname urlprefix\endcsname\relax\def\urlprefix{URL }\fi
\providecommand{\bibinfo}[2]{#2}
\providecommand{\eprint}[2][]{\url{#2}}

\bibitem[{\citenamefont{Kendziorski et~al.}(2005)\citenamefont{Kendziorski,
  Irizarry, Chen, Haag, and Gould}}]{Kendziorski-etal-PNAS2005}
\bibinfo{author}{\bibfnamefont{C.}~\bibnamefont{Kendziorski}},
  \bibinfo{author}{\bibfnamefont{R.~A.} \bibnamefont{Irizarry}},
  \bibinfo{author}{\bibfnamefont{K.~S.} \bibnamefont{Chen}},
  \bibinfo{author}{\bibfnamefont{J.~D.} \bibnamefont{Haag}}, \bibnamefont{and}
  \bibinfo{author}{\bibfnamefont{M.~N.} \bibnamefont{Gould}},
  \bibinfo{journal}{PNAS} \textbf{\bibinfo{volume}{102}}, \bibinfo{pages}{4252}
  (\bibinfo{year}{2005}).

\bibitem[{\citenamefont{Dorfman}(1943)}]{Dorfman1943}
\bibinfo{author}{\bibfnamefont{R.}~\bibnamefont{Dorfman}},
  \bibinfo{journal}{Ann. Math. Stat.} \textbf{\bibinfo{volume}{14}},
  \bibinfo{pages}{436} (\bibinfo{year}{1943}).

\bibitem[{\citenamefont{Gastwirth}(2000)}]{Gastwirth2000}
\bibinfo{author}{\bibfnamefont{J.~L.} \bibnamefont{Gastwirth}},
  \bibinfo{journal}{Am. J. Hum. Genet.} \textbf{\bibinfo{volume}{67}},
  \bibinfo{pages}{1036} (\bibinfo{year}{2000}).

\bibitem[{\citenamefont{Affymetrix}(2004)}]{Affymetrix2004}
\bibinfo{author}{\bibnamefont{Affymetrix}}, \emph{\bibinfo{title}{Sample
  pooling for microarray analysis}} (\bibinfo{year}{2004}),
  \bibinfo{note}{technical note, Affymetrix, San Diego}.

\bibitem[{\citenamefont{Agrawal et~al.}(2002)\citenamefont{Agrawal, Chen, Irby,
  Quackenbush, Chambers, Szabo, Cantor, Coppola, , and
  Yeatman}}]{Agrawal-etal-JNatCancerInst2002}
\bibinfo{author}{\bibfnamefont{D.}~\bibnamefont{Agrawal}},
  \bibinfo{author}{\bibfnamefont{T.}~\bibnamefont{Chen}},
  \bibinfo{author}{\bibfnamefont{R.}~\bibnamefont{Irby}},
  \bibinfo{author}{\bibfnamefont{J.}~\bibnamefont{Quackenbush}},
  \bibinfo{author}{\bibfnamefont{A.~F.} \bibnamefont{Chambers}},
  \bibinfo{author}{\bibfnamefont{M.}~\bibnamefont{Szabo}},
  \bibinfo{author}{\bibfnamefont{A.}~\bibnamefont{Cantor}},
  \bibinfo{author}{\bibfnamefont{D.}~\bibnamefont{Coppola}}, ,
  \bibnamefont{and} \bibinfo{author}{\bibfnamefont{T.~J.}
  \bibnamefont{Yeatman}}, \bibinfo{journal}{J Natl Cancer Inst}
  \textbf{\bibinfo{volume}{94}}, \bibinfo{pages}{513} (\bibinfo{year}{2002}).

\bibitem[{\citenamefont{Churchill and
  Oliver}(2001)}]{Churchill&Oliver-NatureGenetics2001}
\bibinfo{author}{\bibfnamefont{G.~A.} \bibnamefont{Churchill}}
  \bibnamefont{and} \bibinfo{author}{\bibfnamefont{B.}~\bibnamefont{Oliver}},
  \bibinfo{journal}{Nature Genetics} \textbf{\bibinfo{volume}{29}},
  \bibinfo{pages}{355} (\bibinfo{year}{2001}).

\bibitem[{\citenamefont{Jolly et~al.}(2005)\citenamefont{Jolly, Goldstein, Wei,
  Gao, Chen, Huang, Colet, Ryan, Thomas,
  et~al.}}]{Jolly-etal-PhysiolGenomics2005}
\bibinfo{author}{\bibfnamefont{R.~A.} \bibnamefont{Jolly}},
  \bibinfo{author}{\bibfnamefont{K.~M.} \bibnamefont{Goldstein}},
  \bibinfo{author}{\bibfnamefont{T.}~\bibnamefont{Wei}},
  \bibinfo{author}{\bibfnamefont{H.}~\bibnamefont{Gao}},
  \bibinfo{author}{\bibfnamefont{P.}~\bibnamefont{Chen}},
  \bibinfo{author}{\bibfnamefont{S.}~\bibnamefont{Huang}},
  \bibinfo{author}{\bibfnamefont{J.-M.} \bibnamefont{Colet}},
  \bibinfo{author}{\bibfnamefont{T.~P.} \bibnamefont{Ryan}},
  \bibinfo{author}{\bibfnamefont{C.~E.} \bibnamefont{Thomas}}, ,
  \bibnamefont{et~al.}, \bibinfo{journal}{Physiol Genomics}
  (\bibinfo{year}{2005}), \bibinfo{note}{in Press}.

\bibitem[{\citenamefont{Peng et~al.}(2003)\citenamefont{Peng, Wood, Blalock,
  Chen, Landfield, and Stromberg}}]{Peng-etal-BMCBioinformatics2003}
\bibinfo{author}{\bibfnamefont{X.}~\bibnamefont{Peng}},
  \bibinfo{author}{\bibfnamefont{C.~L.} \bibnamefont{Wood}},
  \bibinfo{author}{\bibfnamefont{E.~M.} \bibnamefont{Blalock}},
  \bibinfo{author}{\bibfnamefont{K.~C.} \bibnamefont{Chen}},
  \bibinfo{author}{\bibfnamefont{P.~W.} \bibnamefont{Landfield}},
  \bibnamefont{and} \bibinfo{author}{\bibfnamefont{A.~J.}
  \bibnamefont{Stromberg}}, \bibinfo{journal}{BMC Bioinformatics}
  \textbf{\bibinfo{volume}{4:26}} (\bibinfo{year}{2003}).

\bibitem[{\citenamefont{Shih et~al.}(2004)\citenamefont{Shih, Michalowska,
  Dobbin, Ye, Qiu, , and Green}}]{Shih-etal-Bioinformatics2004}
\bibinfo{author}{\bibfnamefont{J.~H.} \bibnamefont{Shih}},
  \bibinfo{author}{\bibfnamefont{A.~M.} \bibnamefont{Michalowska}},
  \bibinfo{author}{\bibfnamefont{K.}~\bibnamefont{Dobbin}},
  \bibinfo{author}{\bibfnamefont{Y.}~\bibnamefont{Ye}},
  \bibinfo{author}{\bibfnamefont{T.~H.} \bibnamefont{Qiu}}, , \bibnamefont{and}
  \bibinfo{author}{\bibfnamefont{J.~E.} \bibnamefont{Green}},
  \bibinfo{journal}{Bioinformatics} \textbf{\bibinfo{volume}{20}},
  \bibinfo{pages}{3318} (\bibinfo{year}{2004}), \bibinfo{note}{bioinformatics
  Advance Access published on July 9, 2004. doi:10.1093/bioinformatics/bth391}.

\bibitem[{\citenamefont{Kendziorski et~al.}(2003)\citenamefont{Kendziorski,
  Zhang, Lan, and Attie}}]{Kendziorski-etal-Biostatistics2003}
\bibinfo{author}{\bibfnamefont{C.~M.} \bibnamefont{Kendziorski}},
  \bibinfo{author}{\bibfnamefont{Y.}~\bibnamefont{Zhang}},
  \bibinfo{author}{\bibfnamefont{H.}~\bibnamefont{Lan}}, \bibnamefont{and}
  \bibinfo{author}{\bibfnamefont{A.~D.} \bibnamefont{Attie}},
  \bibinfo{journal}{Biostatistics} \textbf{\bibinfo{volume}{4}},
  \bibinfo{pages}{465–477} (\bibinfo{year}{2003}).

\bibitem[{\citenamefont{Jin et~al.}(2001)\citenamefont{Jin, Riley, Wolfinger,
  White, Passador-Gurgel, and Gibson}}]{Jin-etal2001}
\bibinfo{author}{\bibfnamefont{W.}~\bibnamefont{Jin}},
  \bibinfo{author}{\bibfnamefont{R.~M.} \bibnamefont{Riley}},
  \bibinfo{author}{\bibfnamefont{R.~D.} \bibnamefont{Wolfinger}},
  \bibinfo{author}{\bibfnamefont{K.~P.} \bibnamefont{White}},
  \bibinfo{author}{\bibfnamefont{G.}~\bibnamefont{Passador-Gurgel}},
  \bibnamefont{and} \bibinfo{author}{\bibfnamefont{G.}~\bibnamefont{Gibson}},
  \bibinfo{journal}{Nature Genetics} \textbf{\bibinfo{volume}{29}},
  \bibinfo{pages}{389} (\bibinfo{year}{2001}).

\bibitem[{\citenamefont{Muckenthaler et~al.}(2003)\citenamefont{Muckenthaler,
  Roy, Custodio, Minana, deGraaf, Montross, Andrews, and
  Hentze}}]{Muckenthaler-etal2003}
\bibinfo{author}{\bibfnamefont{M.}~\bibnamefont{Muckenthaler}},
  \bibinfo{author}{\bibfnamefont{C.~N.} \bibnamefont{Roy}},
  \bibinfo{author}{\bibfnamefont{A.~O.} \bibnamefont{Custodio}},
  \bibinfo{author}{\bibfnamefont{B.}~\bibnamefont{Minana}},
  \bibinfo{author}{\bibfnamefont{J.}~\bibnamefont{deGraaf}},
  \bibinfo{author}{\bibfnamefont{L.~K.} \bibnamefont{Montross}},
  \bibinfo{author}{\bibfnamefont{N.~C.} \bibnamefont{Andrews}},
  \bibnamefont{and} \bibinfo{author}{\bibfnamefont{M.~W.}
  \bibnamefont{Hentze}}, \bibinfo{journal}{Nature Genetics}
  \textbf{\bibinfo{volume}{34}}, \bibinfo{pages}{102} (\bibinfo{year}{2003}).

\bibitem[{\citenamefont{Zhang and Gant}(2004)}]{Zhang&Gant-Bioinformatics2004}
\bibinfo{author}{\bibfnamefont{S.-D.} \bibnamefont{Zhang}} \bibnamefont{and}
  \bibinfo{author}{\bibfnamefont{T.~W.} \bibnamefont{Gant}},
  \bibinfo{journal}{Bioinformatics} \textbf{\bibinfo{volume}{20}},
  \bibinfo{pages}{2821} (\bibinfo{year}{2004}), \bibinfo{note}{bioinformatics
  Advance Access published on June 4, 2004. Digital Object Identifier (DOI):
  10.1093/bioinformatics/bth336}.

\bibitem[{\citenamefont{Brownlee}(1965)}]{Brownlee1965}
\bibinfo{author}{\bibfnamefont{K.~A.} \bibnamefont{Brownlee}},
  \emph{\bibinfo{title}{Statistical theory and methodology in science and
  engineering}} (\bibinfo{publisher}{John Wiley and Sons, Inc.},
  \bibinfo{year}{1965}).

\bibitem[{\citenamefont{Benjamini and Hochberg}(1995)}]{Benjamini&Hochberg1995}
\bibinfo{author}{\bibfnamefont{Y.}~\bibnamefont{Benjamini}} \bibnamefont{and}
  \bibinfo{author}{\bibfnamefont{Y.}~\bibnamefont{Hochberg}},
  \bibinfo{journal}{J. R. Statist. Soc. B} \textbf{\bibinfo{volume}{57}},
  \bibinfo{pages}{289} (\bibinfo{year}{1995}).

\bibitem[{\citenamefont{Efron}(2004)}]{Efron2004}
\bibinfo{author}{\bibfnamefont{B.}~\bibnamefont{Efron}}, \bibinfo{journal}{J.
  Am. Stat. Assoc.} \textbf{\bibinfo{volume}{99}}, \bibinfo{pages}{96}
  (\bibinfo{year}{2004}).

\bibitem[{\citenamefont{Grant et~al.}(2005)\citenamefont{Grant, Liu, and
  Stoeckert}}]{Grant-etal-Bioinformatics2005}
\bibinfo{author}{\bibfnamefont{G.~R.} \bibnamefont{Grant}},
  \bibinfo{author}{\bibfnamefont{J.}~\bibnamefont{Liu}}, \bibnamefont{and}
  \bibinfo{author}{\bibfnamefont{C.~J.~J.} \bibnamefont{Stoeckert}},
  \bibinfo{journal}{Bioinformatics} \textbf{\bibinfo{volume}{21}},
  \bibinfo{pages}{2684} (\bibinfo{year}{2005}).

\bibitem[{\citenamefont{Storey and
  Tibshirani}(2003)}]{Storey&Tibshrirani-pnas2003}
\bibinfo{author}{\bibfnamefont{J.~D.} \bibnamefont{Storey}} \bibnamefont{and}
  \bibinfo{author}{\bibfnamefont{R.}~\bibnamefont{Tibshirani}},
  \bibinfo{journal}{Proc. Natl. Acad. Sci. USA} \textbf{\bibinfo{volume}{100}},
  \bibinfo{pages}{9440} (\bibinfo{year}{2003}).

\bibitem[{\citenamefont{Pounds and Morris}(2003)}]{Pounds&Morris2003}
\bibinfo{author}{\bibfnamefont{S.}~\bibnamefont{Pounds}} \bibnamefont{and}
  \bibinfo{author}{\bibfnamefont{S.~W.} \bibnamefont{Morris}},
  \bibinfo{journal}{Bioinformatics} \textbf{\bibinfo{volume}{19}},
  \bibinfo{pages}{1236} (\bibinfo{year}{2003}).

\end{thebibliography}
\bibliographystyle{apsrev}

\begin{table}
\caption{ \label{tab-scenario1} For the first scenario described
in the text, the detection power of designs with different levels
of pooling. The $r=1$ panel represents that of non-pooled design.
Other parameter values are: $\sigma ^2=0.05$,
$\sigma_{\epsilon}^2=0.0125$, $\lambda
=\sigma^2/\sigma_{\epsilon}^2=4$, and $m=1$. $N_s=r(n_c+n_t)$ is
the total number of biological subjects required, and
$N_m=m(n_c+n_t)$ is the total number of measurements (microarrays)
needed, counting both the control and the treatment populations.
The preset targets are false positive rate being controlled at
$P_{th}=0.001$, to detect two-fold differential expression
($\mu=1$) with power no less than $0.95$. The minimum number of
biological subjects ($N_s$) and microarrays ($N_m$) that meet the
preset targets are highlighted with bold fonts.}

\begin{tabular}{cccccc}
\hline

$n_c$&$n_t$&$S$&$r$&$N_s$&$N_m$\\

\hline

5&5&0.8175&1&10&10\\
5&6&0.9026&1&11&11\\
6&5&0.9026&1&11&11\\
5&7&0.9488&1&12&12\\
7&5&0.9488&1&12&12\\
6&6&{\bf 0.9553}&1&{\bf 12}&{\bf 12}\\
6&7&0.9796&1&13&13\\
7&6&0.9796&1&13&13\\
\hline

3&3&0.3012&2&12&6\\
3&4&0.5555&2&14&7\\
4&3&0.5555&2&14&7\\
3&5&0.7602&2&16&8\\
5&3&0.7602&2&16&8\\
4&4&0.7937&2&16&8\\
4&5&0.9196&2&18&9\\
5&4&0.9196&2&18&9\\
5&5&{\bf 0.9771}&2&{\bf 20}&{\bf 10}\\
\hline

3&3&0.4060&3&18&6\\
3&4&0.6962&3&21&7\\
4&3&0.6962&3&21&7\\
3&5&0.8774&3&24&8\\
5&3&0.8774&3&24&8\\
4&4&0.9008&3&24&8\\
4&5&{\bf 0.9745}&3&{\bf 27}&{\bf 9}\\
5&4&0.9745&3&27&9\\
5&5&0.9957&3&30&10\\
\hline

2&2&0.0444&5&20&4\\
2&3&0.1930&5&25&5\\
3&2&0.1930&5&25&5\\
2&4&0.4732&5&30&6\\
4&2&0.4732&5&30&6\\
3&3&0.5324&5&30&6\\
3&4&0.8262&5&35&7\\
4&3&0.8262&5&35&7\\
4&4&{\bf 0.9657}&5&{\bf 40}&{\bf 8}\\
\hline

2&2&0.0643&18&72&4\\
2&3&0.2994&18&90&5\\
3&2&0.2994&18&90&5\\
2&4&0.6718&18&108&6\\
4&2&0.6718&18&108&6\\
3&3&0.7309&18&108&6\\
3&4&{\bf 0.9515}&18&{\bf 126}&{\bf 7}\\
4&3&0.9515&18&126&7\\
4&4&0.9969&18&144&8\\

\hline
\end{tabular}
\end{table}

\begin{table}
\caption{\label{tab-exact-vs-approx} Comparison of our exact
results and the approximate results of Ref.
\citep{Shih-etal-Bioinformatics2004}. The upper panel of the table
is for the first scenario, where $\sigma ^2=0.05$,
$\sigma_{\epsilon}^2=0.0125$, $\lambda
=\sigma^2/\sigma_{\epsilon}^2=4$. The lower panel is for the
second scenario, where $\sigma ^2=0.2$,
$\sigma_{\epsilon}^2=0.05$, and $\lambda
=\sigma^2/\sigma_{\epsilon}^2=4$.
 The targets of both scenarios are that the
false positive rate $P_{th}=0.001$ and the power no less than
$S=0.95$. The last column in each panel gives the cost conditions
when pooling samples become beneficial relative to a lower level
of pooling shown in this table.}

\begin{tabular}{cccl}
\hline
$N_m$ & $N_s$ & $N_s$& Conditons\\
      &(Exact)&(Approx)&\\
\hline

11&&11&\\
12&12&&\\
10&20&13&$C_m\ge 4C_s$\\
9&27&16&$C_m\ge 7C_s$\\
8&40&21&$C_m\ge 13C_s$\\
7&126&37&$C_m\ge 86C_s$\\
\hline
30&30&&\\
21&42&&$C_m\ge 1.33C_s$\\
22&&35&\\
16&64&50&$C_m\ge 4.4C_s$\\
14&84&64&$C_m\ge 10C_s$\\
12&180&104&$C_m\ge 48C_s$\\
\hline
\end{tabular}

\end{table}

\end{document}